\documentclass[11pt]{article}

\usepackage{graphicx}                   
\usepackage{amssymb,subfigure}   
\usepackage{verbatim}
\usepackage{color}
\usepackage{xcolor}
\usepackage{amsmath, xparse}

\def\beq{\begin{equation}}
\def\eeq{\end{equation}}
\def\bea{\begin{eqnarray}}
\def\eea{\end{eqnarray}}

\newcommand{\ec}{\end{center}}
\newcommand{\bc}{\begin{center}}

\newcommand{\mb}{\mathbf}

\begin{document}
\begin{center}
{\Large{\bf Strain Fields and Critical Phenomena in Manganites I: Spin-Lattice Hamiltonians}} \\
\ \\
\ \\
by \\
Rohit Singh and Sanjay Puri \\
School of Physical Sciences, Jawaharlal Nehru University, New Delhi -- 110067, India.
\end{center}

\begin{abstract}
We use a model Hamiltonian to study critical phenomena in manganites. This Hamiltonian includes long-range strain interactions, and a coupling between the magnetic order parameter and the strain field. We perform a perturbative renormalization group (RG) analysis and calculate the {\it static} critical exponents, correct to the one-loop level. We compare our RG results with many experiments on doped manganite critical systems. Our theory is in excellent agreement with the experimental values for the critical exponents.
\end{abstract}

\newpage

\section{Introduction}
\label{s1}

Experiments on critical behavior near  paramagnetic-to-ferromagnetic (PM-FM) phase transitions in doped manganites suggest a wide variety of universality classes \cite{mse96,glg98,hkh01,kzh02,nbn03,srg03,pbe07,flh10,ktb12,swo14,mtd15,osh15,tck20}, including tricriticality \cite{krz02,vps08,lpk08,flh10,ntd13}. In some of the samples \cite{nbn03,srg03}, the observed critical behavior is comparable to the universality classes of short-range models (Ising, Heisenberg) of critical phenomena. However, the vast majority of the experimental works \cite{mse96,glg98,kzh02,flh10,ktb12,osh15,tck20} do not fall in the above-mentioned universality classes. Moreover, the exponent values from different experiments are not in agreement, suggesting a range of universality classes. Suzuki \cite{ms74} proposed a scenario where weak universality violation results in a change of the exponents (e.g., spontaneous magnetization exponent $\beta$, susceptibility exponent $\gamma$) which are not directly connected with the correlation length. Universality violation has also been reported numerically in a 2D Ising system with first- and second-neighbour interactions \cite{sq11}. Khan et al. \cite{ksm17} proposed a new scaling theory, which explains universality violation with continuously varying exponents in some doped manganite samples. In the works mentioned above, universality violation arises from modification in the scaling laws. Some earlier works \cite{aa73,ab74,bh76} took into account the effect of additional interactions in the system, leading to tricritical mean-field behavior. Although these works show the existence of a first-order point near the boundary of a second-order line, the diverse critical behavior near tricriticality remains unexplained. Surprisingly, despite the overwhelming experimental evidence in manganites, long-ranged (LR) strain interactions have not received much theoretical attention in formulating a satisfactory theory for describing PM-FM phase transitions.We will address this lacuna in the present paper (I) and its companion paper (II).

Doped manganites R$_{1-x}$A$_x$MnO$_3$, (where $\rm R=La, Ca, Ba, Sr, Pr, Nd$ are trivalent rare earth elements, and $\rm A=Ca, Ba, Sr$ stands for divalent alkaline earth elements) exhibit rich phenomenology on tuning $R$, $A$ and $x$ \cite{xpl20}, e.g., {\it colossal magnetoresistance} (CMR), where a giant increase in electrical resistivity takes place near the Curie temperature \cite{jtm94}. Experiments have shown that a variety of phases, namely, paramagnetic insulating, ferromagnetic metal, antiferromagnet, charged order, orbital order, phase separation, etc., appear in these samples while tuning the temperature ($T$) and the doping concentration ($x$) \cite{sj01,am98}.

As we are interested in formulating an interaction Hamiltonian for such systems, it is relevant to briefly review the physics of manganites. An important mechanism is based on {\it double exchange} (DE), which enables the transfer of an electron between Mn$^{3+}$ to Mn$^{4+}$ ions through an Mn-O-Mn path \cite{cz51,pg60,ko72,nf94,mf01,mf01a}. Although DE is considered an important mechanism to explain the change in resistivity as the system moves from the high-temperature paramagnetic phase to the low-temperature ferromagnetic phase, it underestimates the order of magnetoresistance. In addition, DE does not explain the paramagnetic insulating regime where the electron wave function gets localized. In order to resolve these discrepancies, an additional {\it spin-lattice coupling} has been introduced \cite{mls95,msm96,rzb96}. The origin of this coupling is the buckling of MnO$_6$ octahedra and {\it Jahn-Teller} (JT) lattice distortions. The buckling relates the ionic radius of A-atoms $(r_{\rm A})$ with those of oxygen $(r_{\rm O})$ and manganese $(r_{\rm Mn})$ atoms. This distortion is measured in terms of the tolerance factor:
\beq 
t=\frac{\langle r_{\rm A}\rangle+r_{\rm O}}{{\sqrt 2}(r_{\rm Mn}+r_{\rm O})}.\label{f}
\eeq
By fixing the Mn$^{+3}$/Mn$^{+4}$  ratio and varying the tolerance factor, Hwang et al. \cite{hcr95} showed that the PM-FM critical temperature $T_c$ decreases with decreasing $r_{\rm A}$, and magnetoresistance increases drastically near $T_c$. Further, they also obtained a sharp magnetic transition with decreasing $r_{\rm A}$, indicating a first-order phase transition.

On the other hand, JT lattice distortions are known to be responsible for localizing electrons by forming a self-trapped state termed a {\it polaron}. However, the delocalizing tendencies of electrons compete with localizing effects leading to the polaron wave functions spreading over many lattice sites \cite{am98,ab99,de02,ak02,ssk05,hka06,hks07,ydd10}. This results in the Fr{\"o}lich electron-phonon interaction, given by
\beq
f(r)=\frac{\kappa}{(r^2+1)^{3/2}}e^{-r/\lambda} ,
\eeq
where $\kappa$ is the coupling constant, $r$ is the distance, and $\lambda$ is the range of the interaction. This interaction plays a dominant role compared to magnetic exchange interactions \cite{ab99,de02,ssk05} in such systems. Using the quantum Monte Carlo approach \cite{hka06}, it has been shown that the lattice coordination number becomes unimportant as $\lambda$ is increased. In that case, the physical properties of manganite systems are primarily determined by the dimensionality.

Apart from the above theoretical works, several experiments \cite{hcr95,dzm96,sxk96,zck96,bby98,bbk98,am98,yza06,cty06,xbh08,onw09,whl10,ykb10} and other theoretical works \cite{mdm98,am01,kk03,als03,alb04,als05,dyz10,lr10,bbc11} have also highlighted the role of strain fields in determining the critical properties of manganites. An extreme sensitivity of $T_c$ to biaxial strain was predicted by Millis et al. \cite{mdm98}. Khomskii and Kugel \cite{kk03} showed the formation of various charge and orbital ordered structures due to strain fields. The generalization of atomic-scale spin-lattice distortions to continuum elasticity helped to understand texturing and domain wall profiles in such compounds \cite{als03,alb04,als05}. The above works suggested the importance of LR strain modes and spin-lattice coupling to understand critical phenomena in these strongly-correlated systems.

These two papers (I and II) report a comprehensive study of strain effects on critical phenomena in manganites. We undertake a renormalization group (RG) study of an appropriate coarse-grained Hamiltonian. A novel feature of our study is a detailed comparison of the RG critical exponents with experimental results available in the literature. For this purpose, we treat the decay exponent of the LR interactions $\sigma$ as an adjustable parameter. In all cases, the agreement with experiments is excellent.

There have been some earlier theoretical studies of the effect of elastic degrees of freedom near a critical point \cite{br62,hw70,ws70,aa73,ab74,bh76}. Some of these studies  \cite{ws70,aa73,ab74} incorporated an LR interaction term in a Ginzburg-Landau Hamiltonian. An RG analysis of this Hamiltonian \cite{bh76,sdn15} shows a first-order transition with tricritical mean-field exponents \cite{sm76}. These studies suggest two crucial facts: \\
(a) the spin-lattice interaction plays an essential role in manganites, \\
(b) the effective Hamiltonian of such a system must be nonlocal due to the LR character of the strain modes, and it may result in a first-order transition with tricritical mean-field behavior.

With this background, we study the following model Hamiltonian in this paper:
\bea
\mathcal H[\Phi,\psi] &=& \int d^d\mb x \bigg[\frac{r_0}{2}\Phi^2(\mb x) + \displaystyle\frac{c_0}{2} |\nabla\Phi(\mb x)|^2+\frac{\kappa_0^{-1}}{2} \psi(\mb x)(m^2-\nabla^2)^{\sigma}\psi(\mb x)+ \nonumber \\
&& g_0\psi(\mb x)\Phi^2(\mb x) \bigg].
\label{H}
\eea
Here, $\Phi$ is the $n$-component magnetic order parameter, and $\psi$ is the strain field. In Eq.~(\ref{H}), the parameters $r_0$ and $c_0$, have their usual interpretation; and $\kappa_0^{-1}$ is the interaction strength of the LR strain-strain term with range $m^{-1}$ and exponent $\sigma$. The LR term may also be written in the nonlocal form $\int d^d \mb x~\int d^d \mb x'~\psi(\mb x)~u (\mb x - \mb x')~\psi (\mb x')$, where $u(\mb y) \sim e^{-my}/y^{d+\sigma}$. The final term on the right-hand-side models the spin-lattice interaction of strength $g_0$.

There have been several earlier studies of models with nonlocal and non-analytic interactions. In the context of $\Phi^4$ theory, such terms have been considered as possible generalizations of the quadratic ($\Phi^2$) term \cite{fmn72,apr14}. In recent work, Defenu et al. \cite{dtc15,dtr17} have studied nonlocal spin-spin interactions in the context of long-range interacting systems in cold atom applications. The phase diagram shows a non-trivial dependence on $\sigma$, and the authors found multi-critical universality classes. Such interactions have also been found to be useful to understand stripe formation in dipolar magnetic films near their classical and quantum phase transition points \cite{msn15,mbs17}. A common feature of the above works is the incorporation of the LR coupling in the $\Phi^2$-term, while the interaction term $\Phi^4$ remains short-ranged. Our approach in this paper is analogous to the LR generalization of the $\Phi^{4}$-term. We stress that both classes of theories (i.e., non-local $\Phi^2$ vs. non-local $\Phi^4$) yield a continuously varying range of critical exponents.

Before proceeding, it is also relevant to contrast our present approach with earlier works. Bergman and Halperin \cite{bh76} studied the effect of a compressible lattice on critical phenomena and found a signature of first-order critical behavior. The usual Ginzburg-Landau model requires a $\phi^6$-term to explain a tricritical point \cite{sm76}. However, a mode-coupling interaction results in a tricritical point without such additional terms \cite{sdn15}. In this work, we explicitly incorporate the Frolich-type electron-phonon interaction to study critical phenomena in manganites. As mentioned above, for large screening length, the observed physical behavior of such systems is governed by dimensionality \cite{hka06}. Our RG calculations support this argument as the observed critical behavior of our model depends on dimensionality $d$ and $\sigma$. In particular, we demonstrate that incorporation of such interactions enables us to capture a wide variety of available experimental exponents in manganites -- both near and far from tricriticality.

We thus study the critical behavior of the Hamiltonian in  Eq.~(\ref{H}) via RG analysis in the limit of wave-vector $k\rightarrow 0$ and $ m \rightarrow0$. Our limiting procedure will be $k\rightarrow 0$, followed by $m\rightarrow 0$ (see Ref.~\cite{fw71} for a detailed discussion). In this paper (I), we restrict ourselves to studying static aspects of PM-FM phase transitions in manganite systems. A generalization of the above model is studied in II to understand the effect of strain fields on critical behavior at the paramagnetic-to-antiferromagnetic (PM-AFM) transition. In II, we obtain both static and dynamical critical exponents in the framework of {\it Model C} \cite{hh77}.

This paper is organized as follows. In Sec.~\ref{s2}, we present detailed RG calculations at one-loop order. We find the non-trivial fixed point and derive closed-form expressions for the critical exponents. In Sec.~\ref{s3}, we compare our calculated critical exponents with the available experimental results for doped manganite systems. Finally, we conclude this paper with a summary and discussion in Sec.~\ref{s4}.

\section{Perturbative Renormalization Group Analysis}
\label{s2}

In this section, we present the RG analysis of the Hamiltonian in Eq.~(\ref{H}). Calculating the self-energy and vertex corrections at one-loop for the  parameters $r_0$, $c_0$, and $g_0$, we subsequently construct the RG recursion relations. From these recursion relations, we identify the non-trivial fixed point and obtain the critical exponents. The details of our calculation are presented below.

\subsection{Momentum Shell Decimation}
\label{s21}

Here, we present our perturbative RG scheme at one-loop order and calculate various self-energy and vertex corrections. To do so, we write the $d$-dimensional Fourier transform for the $n$-component vector order parameter $\Phi$ (with components $\phi_i$) and the strain field $\psi$:
\beq
f(\mb x)=\int \frac{d^{d}k}{(2\pi)^d}f(\mb k)e^{i\mb k\cdot\mb x},
\eeq
where $f$ is either $\phi_i$ or $\psi$. We write the Hamiltonian in Eq.~(\ref{H}) in Fourier space as
\bea
\mathcal H &=& \sum_{i=1}^{n}\int_{0}^{\Lambda}\frac{d^d
k}{(2\pi)^{d}}\frac{r_0+c_0k^{2}}{2}|\phi_{i}(\mb
k)|^{2} + \frac{\kappa_0^{-1}}{2}\int_{0}^{\Lambda}\frac{d^d k}{(2\pi)^{d}}(k^2+m^2)^{\sigma}|\psi(\mathbf k)|^{2} \nonumber \\
&& + g_0\sum_{i=1}^{n}\int_{0}^{\Lambda}\int_{0}^{\Lambda}\frac{d^{d} k_1}{(2\pi)^{d}}\frac{d^{d} k_{2}}{(2\pi)^{d}}\psi(\mb k_{1})\phi_{i}(\mb k_{2})\phi_{i}(-\mb k_1-\mb k_2) .
\label{HF}
\eea
The above Fourier-transformed Hamiltonian contains all modes less than the cut-off $\Lambda \sim a^{-1}$, where $a$ is the microscopic parameter (lattice spacing) of the theory. The inclusion of $\Lambda$ avoids ultraviolet divergences, and the mode elimination process
begins from $k\sim \Lambda$. In Eq.~(\ref{HF}), the momentum integrations lie between small $k\sim0$ to large $k\sim \Lambda$ values. The components with $0 \leqslant k \leqslant \Lambda/b$ ($b>1$) are termed {\it slow modes}, while components with $\Lambda/b \leqslant k \leqslant \Lambda$ are called {\it fast modes}. Therefore, the large wavelength limit ($k\sim0$)  is analyzed in the momentum shell decimation context by eliminating small wavelengths ($k\sim \Lambda$) through the high momentum end. After successive mode elimination, one achieves the large-scale behavior. 

The two-point bare correlation function of the order parameter $\phi_i$ and the scalar field $\psi$ can be expressed as \cite{sm76}
\bea
\langle \phi_i ({\mathbf k}) \phi_j ({\mathbf k'}) \rangle_0 &=& G_0({\mathbf k})(2\pi)^{d}\delta^{d}({\mathbf k}+{\mathbf k'}) \delta_{ij} , \\ \label{correlation1}
\langle\psi({\mathbf k}) \psi({\mathbf k'})\rangle_0 &=& D_0({\mb k})(2\pi)^{d}\delta^{d}({\mathbf k}+{\mathbf k'}) .
\label{correlation2}
\eea
Here, $\langle..\rangle_0$ indicates two-point expectation values with respect to the Gaussian part of the Hamiltonian in Eq.~(\ref{HF}). We define
\beq G_0(\mathbf k)=(r_{0}+c_0\mathbf k^{2})^{-1} 
\eeq
and 
\beq
D_0(\mb k)=\kappa_0(\mathbf k^{2}+m^2)^{-\sigma}
\eeq
as the bare propagators for $\phi_i$ and  $\psi$ fields, respectively. With this, we carry out an RG analysis via Wilson's momentum shell decimation scheme \cite{wk74,sm76} in one-loop order.  The procedure begins with the  elimination of modes $\phi_i^{>}(\mb k)$ and $\psi^{>}(\mb k)$ lying in the momentum range $\Lambda/b\leqslant k\leqslant\Lambda$. This transforms the Hamiltonian in terms of the remaining modes $\phi_i^{<}(\mb k)$, $\psi^{<}(\mb k)$ in the reduced range $0\leqslant k\leqslant \Lambda/b$. As we show, this process yields the renormalized corrections to the bare parameters $r_0$, $c_0$, and $g_0$ at one-loop order of the perturbation expansion.

\begin{figure}
\begin{center}
\includegraphics[width=0.6\textwidth,height=!]{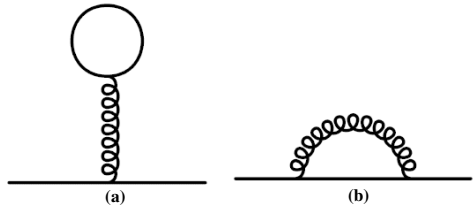}
\caption{Feynman diagrams giving self-energy corrections to $r_0$ and $c_0$. The internal straight lines represent the correlation between the fast modes of the $\Phi$-field, and the wiggly lines represent the correlation between the fast modes of the $\psi$-field.}
\label{f1}
\end{center}
\end{figure}

The relevant Feynman diagrams are  given in Fig.~\ref{f1}, and yield the self-energy corrections to the bare parameters $r_0$ and $c_0$:
\bea
\Sigma_a(\mb 0) &=& -ng_0^2\int_{\Lambda/b}^{\Lambda}\frac{d^d q}{(2\pi)^{d}}D_0^{>}(\mb 0)G_0^{>}(\mb q) , \\
\label{sigmaa}
\Sigma_b(\mb k) &=& -2g_0^2\int_{\Lambda/b}^{\Lambda}\frac{d^d q}{(2\pi)^{d}}D_0^{>}(-\mb q-\mb k)G_0^{>}(\mb q).
\label{sigmab}
\eea
The above self-energy corrections are obtained from the amputated part of diagrams in Figs.~\ref{f1}(a) and \ref{f1}(b). In Fig.~\ref{f1}, gluon lines represent the correlation between the fast modes of strain fields $\psi^{>}$ and the internal solid lines in the loop come from the contraction of fast modes of $\phi_{i}^{>}$. We point out that Eq.~(\ref{sigmaa}) generates a non-analytic contribution without the screening parameter $m$ in the $k\rightarrow 0$ limit for $\sigma > 0$. However, we finally set $m \rightarrow 0$ to see the critical behavior of the above model (see Ref.~\cite{fw71} for a detailed discussion). In order to evaluate the integrals in the range $\Lambda/b\leqslant q\leqslant \Lambda$, we consider  the limit $q\gg m$ \cite{fw71} which, in turn, indicates that $m^{-1}\gg a$. 
This also implies that the range of interaction is much larger than the lattice spacing. Since in the momentum shell decimation scheme, the effect of elimination of small scales (internal momentum $q$) on the large scales (external momenta $k$) is to be calculated. The above corrections are expanded in the limit $q \gg k$, which is equivalent to an expansion about $k= 0$. To capture the large-scale behavior, we consider the following expansion in the above expressions for $\Sigma_a$ and $\Sigma_b$:
\bea
D_0(-\mb q-\mb k) &=& \frac{\kappa_0}{{\left[(-\mathbf k-\mb q)^{2}+m^{2}\right]}^{\sigma}} \nonumber \\
&=& \frac{\kappa_0}{q^{2\sigma}}\left[1-2\sigma\frac{\mb k\cdot\mb q}{q^2}-\sigma\frac{k^2+m^2}{q^2}+2\sigma(\sigma+1)\frac{(\mb k\cdot \mb q)^2}{q^4}+\ldots\right]
\label{d}.
\eea

Thus, we obtain the total correction to $r_0$ and $c_0$ as 
\beq
\Sigma_a(\mb 0)+\Sigma_b(\mb k)=\frac{1}{2}\Delta r+\frac{1}{2}\Delta c\,k^2+\ldots,
\label{Sigma_exp}
\eeq
where
\bea
\Delta r &=& -\frac{2ng_0^2 \kappa_0S_{d}}{m^{2\sigma}(2\pi)^{d}}\left[\frac{(b^{2-d}-1)\Lambda ^{d-2}}{c_0(2-d)}-\frac{r_{0}}{c_0^{2}}\frac{(b^{4-d}-1)\Lambda^{d-4}}{(4-d)}\right] - \nonumber \\
&& \frac{4g_0^2\kappa_0S_d}{(2\pi)^{d}} \left[\frac{(b^{2-d+2\sigma}-1)\Lambda ^{d-2-2\sigma}}{c_0(2-d+2\sigma)}-\left(\frac{r_0}{c_0^2}+\frac{\sigma m^2}{c_0} \right) \frac{(b^{4-d+2\sigma}-1)\Lambda^{d-4-2\sigma}}{(4-d+2\sigma)} \right], 
\label{delr} \nonumber \\
\eea
and
\beq
\Delta c=-\frac{4g_0^2\kappa_0S_d}{c_0(2\pi)^d} \frac{\sigma(2\sigma+2-d)}{d}\frac{(b^{4-d+2\sigma}-1)\Lambda ^{d-4-2\sigma}}{(4-d+2\sigma)} .
\label{delc}
\eeq
In Eqs.~(\ref{delr})-(\ref{delc}), $S_d= 2\pi^{d/2}/\Gamma(d/2)$ is the surface area of a unit sphere in $d$ space dimensions. We see that the parameter $\kappa_0^{-1}$ does not acquire any correction at this order of calculation ($\Delta\kappa^{-1}=0$).

\begin{figure}
\begin{center}
\noindent\includegraphics[width=0.5\textwidth,height=!]{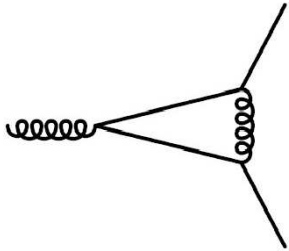}
\caption{Feynman diagram for correction to the bare vertex $g_0$. The straight and wiggly lines have the same meaning as in Fig.~\ref{f1}.}
\label{f2}
\end{center}
\end{figure}

We also obtain the correction to the interaction vertex $g_0$ from the  Feynman diagram given in Fig.~\ref{f2} which is expressed as
\beq
\Pi(\mb k_1, \mb k_2)=4g_0^{3}\int_{\Lambda/b}^{\Lambda}\frac{d^d q}{(2\pi)^{d}}D_0^{>}(\mb q-\mb k_2)G_0^{>}(\mb q)G_0^{>}(-\mb k_1-\mb q).\label{Pi}
\eeq 
In the above expression, the momentum integration lies in the high-momentum shell $\Lambda/b\leqslant q\leqslant\Lambda$. Using Eq.~(\ref{d}) in the limit of vanishing external momentum $k_1$ and $k_2$, the integral $\Pi(\mb k_1, \mb k_2)$ contributes a correction to the bare coupling constant $g_0$. Solving  this integral in the limits $q\gg k_i (i=1,2)$ and  $q\gg m$, Eq.~(\ref{Pi}) yields
\beq
\Delta g=\frac{4g_0^3 \kappa_0S_d}{(2\pi)^d}\left[\frac{(b^{4+2\sigma-d}-1)\Lambda^{d-4-2\sigma}}{c_0^2(4+2\sigma-d)}-\left(\frac{\sigma m^{2}}{c_0^2}+\frac{2r_{0}}{c_0^3}\right)\frac{(b^{6-d+2\sigma}-1)\Lambda^{d-2\sigma-6}}{(6-d+2\sigma)}\right]
\label{delgamma}.
\eeq 
 
\subsection{RG Flow and Static Critical Exponents}
\label{s22}

The momentum shell decimation scheme integrates  the modes lying in the range  $\Lambda/b \leqslant k \leqslant \Lambda$ and lowers the cut-off. To restore the cut-off, we rescale momenta and fields as 
\bea
\mb k' &=& {b\mathbf k}, \nonumber \\
\Phi'(\mb k') &=& b^{-x}\Phi^{<}(\mb k), \nonumber \\
\psi'(\mb k) &=& b^{y}\psi^{<}(k) .
\eea
The exponents $x$ and $y$ are obtained from the two-point spin-spin correlation [$\sim r^{-(d-2+\eta)}$] and strain-strain correlation [$\sim r^{-(d-h)}$] at the transition point as
\bea
x &=& \frac{1}{2} (d+2-\eta) , \nonumber \\
y &=& -\frac{1}{2}(d+h) .
\label{zeta}
\eea
The exponents $\eta$ and $h$ will be determined from the  RG flow equations given below. We thus write the RG recursion relations for the model parameters $r_0$, $c_0$, $\kappa_0^{-1}$, and $g_0$ as
\bea
r' &=& b^{2-\eta}(r_{0}+\Delta r), \label{recursion1} \\
c' &=& b^{-\eta}(c_0+\Delta c), \label{recursion2} \\
(\kappa^{-1})' &=& b^{h-2\sigma}(\kappa_0^{-1}+\Delta \kappa) \label{recursion3} \\
g' &=& b^{4-d-2\eta+h}(g_0+\Delta g). \label{recursion3}
\eea

We introduce $b=e^{\delta l}$ and take the limit $\delta l\rightarrow0$ to arrive at the RG flow equations for the scale dependent parameters $r(l)$, $c(l)$, $\kappa^{-1}(l)$ and $g(l)$:
\bea
&& \frac{dr}{dl} = (2-\eta)r-\frac{2ng^2 \kappa S_d}{m^{2\sigma} (2\pi)^d}\left(\frac{\Lambda^{d-2}}{c}-\frac{r}{c^2}\Lambda^{d-4}\right) \nonumber \\
&& \quad \quad - \frac{4g^2\kappa S_d}{c(2\pi)^d}\left[\frac{\Lambda^{d-2-2\sigma}}{c}-\left(\frac{r}{c^2}+\frac{\sigma m^2}{c}\right)\Lambda^{d-4-2\sigma}\right], \label{flowr0} \\
&& \frac{dc}{dl} = -\eta c-\frac{4g^2\kappa \sigma(2\sigma+2-d)S_d}{d (2\pi)^d}\frac{\Lambda^{d-4-2\sigma}}{c},
\label{flowc} \\
&& \frac{d\kappa^{-1}}{dl} = (h-2\sigma)\kappa^{-1}
\label{flowC}, \\
&& \frac{dg}{dl}=\frac{(4-d-2\eta+h)g}{2}+\frac{4g^3\kappa S_d}{c^2(2\pi)^d}\left[\frac{\Lambda^{d-4-2\sigma}}{c^2}-\left(\frac{2r}{c^3}+\frac{\sigma m^2}{c^2}\right)\Lambda^{d-6-2\sigma} \right] \label{flowlambda}. \nonumber \\
\eea
The above flow equations depend upon the dimension $d$, long-range exponent $\sigma$, and the number of components $n$ of the order parameter. We notice a dependence on the coupling parameter $g^2 \kappa$. Thus, it is reasonable to define a new coupling constant $u=-g^2 \kappa/2$. The corresponding flow equations for $r, c, u$ are
\bea
&& \frac{dr}{dl}=(2-\eta)r+\frac{4nu S_d}{m^{2\sigma} (2\pi)^d}\left(\frac{\Lambda^{d-2}}{c}-\frac{r}{c^2}\Lambda^{d-4}\right)+
\nonumber \\
&& \quad \quad \quad \frac{8u S_d}{(2\pi)^d}\left[\frac{\Lambda^{d-2-2\sigma}}{c}-\left(\frac{r}{c^2}+\frac{\sigma m^2}{c}\right)\Lambda^{d-4-2\sigma}\right], \label{flowr0} \\
&& \frac{dc}{dl} = -\eta c+\frac{8u\sigma(2\sigma+2-d)S_d}{d (2\pi)^d}\frac{\Lambda^{d-4-2\sigma}}{c}, \label{flowc} \\
&& \frac{du}{dl}=(4-d-2\eta+2\sigma)u-\frac{16 u^2 S_d}{(2\pi)^d}\left[\frac{\Lambda^{d-4-2\sigma}}{c^2}-\left(\frac{2r}{c^3}+\frac{\sigma m^2}{c^2}\right)\Lambda^{d-6-2\sigma}\right] \label{flowlambda}.
\nonumber \\
\eea
We see that the parameter $c$ does not flow to a fixed point. However, the renormalization of $c$ leads to a finite Fisher exponent $\eta$.
Using the above flow equations, we obtain the non-trivial fixed point in $(r,u)$-space as
\beq
\frac{r^{*}}{c}=-\frac{(4-d-2\eta+2\sigma)\left[\frac{2n}{w^{\sigma}}+{4}(1-\sigma w)\right]\Lambda^2}{(2-\eta)\left[8(1-\sigma w)\right]-(4-d-2\eta+2\sigma)(\frac{2n}{w^\sigma}+4)},
\label{fpr0}
\eeq
and
\beq
\frac{u^*}{c^{2}}=\frac{(4-d-2\eta+2\sigma)\Lambda^{4-d+2\sigma}}{\frac{16S_d}{(2\pi)^d}(1-\sigma w)},
\label{fplambda}
\eeq
where $w=m^2/\Lambda^2$ is a dimensionless screening parameter.

Introducing slight deviations $\delta r=r-r^*$ and $\delta u=u-u^*$ around the fixed point, a linear stability analysis yields the matrix equation:
\beq
\left[ \begin{array}{c}d(\delta r)/dl\\ d(\delta u)/dl \end{array}\right] = \begin{bmatrix} \lambda_1 & \frac{8S_d}{(2\pi)^d}\frac{\Lambda^{d-2\sigma-2}}{c}\left(1-\sigma w-\frac{r^*}{c\Lambda^2}\right)+\frac{4n}{w^\sigma}\frac{8S_d}{(2\pi)^d}\frac{\Lambda^{d-2\sigma-2}}{c}\\ 0+O(u^2) & \lambda_2\end{bmatrix}\left[\begin{array}{c}\delta r\\\delta u \end{array}\right] .
\eeq
The eigenvalues $\lambda_1$ and $\lambda_2$ correspond to unstable and stable eigen directions, respectively, in the RG flow. We find
\bea
&& \lambda_1=2-\eta-(4-d-2\eta+2\sigma) \frac{(\frac{n}{w^{\sigma}}+2)}{4(1-\sigma w)}, \nonumber \\
&& \lambda_2=d-4+2\eta-2\sigma .
\label{eig}
\eea 
Further, the marginal stability of the stable eigenvalue $\lambda_2$ gives  the upper critical dimension $d_c$ as $d_c=4+2\sigma$. 

Finally, we calculate the critical exponents in an $\epsilon=d_c-d=4+2\sigma-d$ expansion scheme. Using the fixed point value  $u^*$ in Eq.~(\ref{flowc}), we obtain
\beq
\eta=-\frac{\sigma\epsilon}{4-4\sigma w-2\sigma^2w}+O(\epsilon^2) . \label{eta}
\eeq
The correlation length exponent $\nu$ is calculated from $\lambda_1$ as $\nu=1/\lambda_1$, giving
\beq
\nu=\frac{1}{2}-\frac{\sigma\epsilon}{4(4-4\sigma w-2\sigma^2w)}+\frac{\epsilon\left(\frac{n}{w^\sigma}+2\right)}{4(1-\sigma w)}\left[\frac{\sigma}{2(4-4\sigma w-2\sigma^2 w)}+\frac{1}{4}\right]+O(\epsilon^2).
\eeq
The other  critical exponents, e.g., specific heat exponent $\alpha$, magnetization exponent $\beta$, susceptibility exponent $\gamma$, and the critical isotherm exponent $\delta$ can be  derived from $\eta$ and $\nu$ using known scaling relations, namely, {\it Josephson}: $\nu d=2-\alpha$, {\it Fisher}: $\gamma=\nu(2-\eta)$, {\it Widom}: $\gamma=\beta(\delta-1)$, and {\it Rushbrooke}: $\alpha+2\beta+\gamma=2$. They are obtained as
\bea
&& \alpha=-\sigma+\epsilon\left[\frac{1}{2}+\frac{\sigma(\sigma+2)}{2(4-4\sigma-2\sigma^2 w)}\left\{1-\frac{(\frac{n}{w^{\sigma}}+2)}{2(1-\sigma w)}\right\}-\frac{(\sigma+2)(\frac{n}{w^\sigma}+2)}{8(1-\sigma w)} \right]+O(\epsilon^2), \label{alpha} \nonumber \\ \\
&& \beta=\frac{(\sigma+1)}{2}-\frac{\epsilon}{4}\left[1+\frac{1}{4-4\sigma w-2\sigma^2w}\left\{\sigma(\sigma+2)-\frac{\sigma(\sigma+1)(\frac{n}{w^\sigma}+2)}{2(1-\sigma w)}\right\} \right. - \nonumber \\
&& \quad \quad \quad \left. \frac{(\sigma+1)(\frac{n}{w^\sigma}+2)}{4(1-\sigma w)} \right]+O(\epsilon^2), \label{beta} \\
&& \gamma=1+\frac{\epsilon\left(\frac{n}{w^\sigma}+2\right)}{4(1-\sigma w)}\left[\frac{1}{2}+\frac{\sigma}{4-4\sigma w-2\sigma^2w}\right]+O(\epsilon^2),
\label{gamma} \\
&& \delta=\frac{\sigma+3}{\sigma+1}+\frac{\epsilon}{(\sigma+1)^2}\left[1+\displaystyle\frac{\sigma(\sigma+2)}{4-4\sigma w-2\sigma^2 w}\right]+O(\epsilon^2).
\label{delta}
\eea

The above exponents depend on the dimensionless parameter $w$, so they may be considered non-universal. An analogous situation arises in the case of the Ashkin-Teller Potts model which yields critical exponents varying continuously with the parameter occurring in the four-spin term of the model Hamiltonian \cite{lk79}. Furthermore, as we are interested in large-scale interactions near the critical point, we analyze the limiting behavior $w\rightarrow 0$ in Eqs.~(\ref{eta})--(\ref{delta}) and arrive at the critical exponents 
\bea
&& \eta=-\frac{\sigma\epsilon}{4}+O(\epsilon^2), \label{etal} \\
&& \nu=\frac{1}{2}+\frac{\epsilon}{8}+O(\epsilon^2), \\
&& \alpha=-\sigma\left(1+\frac{\epsilon}{4}\right)+O(\epsilon^2), \label{alphal} \\
&& \beta=\frac{1+\sigma}{2}-\frac{\epsilon(2-\sigma)}{16}+O(\epsilon^2), \label{betal} \\
&& \gamma=1+\displaystyle\frac{\epsilon}{4}\left(1+\frac{\sigma}{2}\right) +O(\epsilon^2), \label{gammal} \\
&& \delta=\frac{3+\sigma}{1+\sigma}+\frac{\epsilon}{(1+\sigma)^2}\left[1+\displaystyle\frac{\sigma(\sigma+2)}{4} \right]+O(\epsilon^2). \label{deltal}
\eea   

The above calculated critical exponents depend on the dimension $d$, and the LR exponent $\sigma$. For the critical fixed point, we need stable and unstable eigenvalues along the $r$ and $u$ axis, respectively. This is equivalent to the condition that $\lambda_1>0$ and $\lambda_2 < 0$ in Eq.~(\ref{eig}) \cite{sm76}. Thus, the allowed range of $\sigma$ depends upon $d$. We find $-1/2<\sigma<0$ in $d=3$, and $-1<\sigma<0$ in $d=2$. The mean-field limit ($\epsilon = 0$) of Eqs.~(\ref{etal})-(\ref{deltal}) is an example of one of the three scenarios discussed in Ref.~\cite{ksm17} where $\delta$ changes and $\gamma$ is fixed. For $d < d_c=4+2\sigma$, we see the continuous variation of both $\delta$ and $\gamma$, and the theory yields complete non-universal behavior. In the next section, we compare the RG critical exponents with experimental results for manganite systems.

\section{Comparison with Experiments}
\label{s3}

In the doped manganites (R$_{1-x}$A$_x$MnO$_3$), different choice of R, A, and $x$ lead to different critical exponents near and away from tricriticality as evident in Table~\ref{comp}. For example, in polycrystalline La$_{0.6}$Ca$_{0.4}$MnO$_3$ \cite{krz02}, a {\it modified Arrot plot} (MAP) and {\it critical isotherm} (CI) analysis yield $\alpha=0.48\pm0.06$, $\beta=0.25\pm 0.03$, $\gamma=1.03 \pm 0.05$, and $\delta=5.0\pm 0.8$. In polycrystalline La$_{0.1}$Nd$_{0.6}$Sr$_{0.3}$MnO$_3$ \cite{flh10}, measurements with MAP and CI analysis yield $\beta=0.248\pm 0.006$, $\gamma=1.066\pm 0.002$, and $\delta=5.17\pm 0.02$. In polycrystalline Nd$_{0.67}$Sr$_{0.33}$MnO$_3$ \cite{vps08}, the magnetization data analyzed with MAP yields $\beta=0.23\pm 0.02$, $\gamma=1.05 \pm 0.03$, and $\delta=5.13 \pm 0.04$. The above critical exponents are close to those in tricritical mean-field theory ($\beta=1/4,\gamma=1$, and $\delta=5$) \cite{sm76}. However, there are other  compounds \cite{mse96,glg98,hkh01,kzh02,nbn03,srg03,pbe07,flh10,ktb12,swo14,mtd15,osh15,tck20} whose  critical exponents differ from the tricritical mean-field values. For instance, in single crystal La$_{0.7}$Sr$_{0.3}$MnO$_3$,  the critical exponents obtained via MAP and CI analysis are $\beta=0.37\pm0.04,\gamma=1.22\pm0.03$, and $\delta=4.25\pm0.2$ \cite{glg98}. In polycrystalline  $\mbox{Pr}_{0.6}\mbox{Sr}_{0.4}\mbox{MnO}_3$, MAP and CI analysis yields  $\beta=0.314 \pm0.0006, \gamma=1.095\pm0.007$, and $\delta=4.545\pm0.008$ \cite{osh15}.

There are also several works where a change in $x$ in the same material leads to variation in the critical exponents \cite{glg98,hkh01,krz02,nbn03,vps08}. For example, different critical indices were reported for La$_{1-x}$Ca$_{x}$MnO$_3$ when $x=0.2$ \cite{hkh01} and $x=0.4$ \cite{krz02}. Similar behavior was found in La$_{1-x}$Sr$_x$MnO$_3$ when $x=0.3$ \cite{glg98} and $x=0.125$ \cite{nbn03}. This was also observed in Ref.~\cite{vps08} for a different compound, namely, Nd$_{1-x}$Sr$_x$MnO$_3$ with $x=0.33$ and $x=0.4$. 

In Table~\ref{comp}, we compare our RG values of $\beta, \gamma$ and $\delta$ at $O(\epsilon)$ to those from experiments on doped manganites. For comparison, we first match the $\beta$-values (up to three significant figures) from Eq.~(\ref{betal}) to determine the appropriate $\sigma$ in the allowed range. Using this value of $\sigma$, we obtain $\gamma$ and $\delta$ from Eqs.~(\ref{gammal})-(\ref{deltal}).

The observed values of $\beta$ in Refs.~\cite{krz02,vps08,flh10,ntd13} are $0.25\pm0.03$, $0.248$, $0.23\pm 0.02$, and $0.248\pm0.006$, respectively. The RG calculation requires $-1/2<\sigma<0$ in $d=3$, so that $0.250< \beta <0.375$. Thus, the experiments with $\beta=0.23\pm0.02$ and $\beta=0.248\pm0.006$ cannot be matched (without error bars) with our RG values of $\beta$. At $\sigma=-0.5$, the theory becomes marginal through a crossover to the Gaussian fixed point. Thus, the closest $\sigma$-value to model these experiments is chosen to be $\sigma=-0.499$.
From Eqs.~(\ref{betal})-(\ref{deltal}), $\sigma=-0.5$ gives $\epsilon = 0$ in $d=3$. This yields the tricritical mean-field exponents $\beta=1/4, \gamma=1$, and $\delta=5$. We stress that the current theory invokes an expansion about tricritical mean-field theory, contrary to the conventional $\Phi^4$-theory and the long-range model of Fisher et al. \cite{fmn72}.
The experimental results \cite{mse96,glg98,hkh01,kzh02,nbn03,srg03,pbe07,flh10,ktb12,swo14,mtd15,osh15,tck20} which deviate from tricritical mean-field values are also obtained from the present theory by $\sigma$ in the permitted range, as shown in Table~\ref{comp}. For example, in single crystal La$_{0.7}$Sr$_{0.3}$MnO$_3$ \cite{glg98}, the experimentally observed critical exponents  $\beta=0.37\pm0.04$,  $\gamma=1.22\pm0.03$, and $\delta=4.25\pm0.2$ are captured for $\sigma=-0.015$. In polycrystalline $\mbox{Pr}_{0.6}\mbox{Sr}_{0.4}\mbox{MnO}_3$ \cite{osh15}, the exponents $\beta=0.314 \pm0.0006, \gamma=1.095\pm0.007$, and $\delta=4.545\pm0.008$ arise for $\sigma=-0.213$. 

Our model Hamiltonian matches the experimental results for $\sigma$ lying in the range $-0.5 < \sigma < 0$. In some experiments \cite{srg03,pbe07}, the observed values of $\beta$, $\gamma$ and $\delta$ are close to those of short-ranged models (Ising, Heisenberg). However, the vast majority of samples \cite{mse96,glg98,hkh01,kzh02,flh10,ktb12,swo14,mtd15,osh15} in Table~\ref{comp} show behavior away from those theories. Further, RG analysis of the LR model of Fisher et al. \cite{fmn72} produces exponents near mean-field values ($\beta=0.5$, $\gamma=1.0$, $\delta=3.0$). Thus, this model is unable to predict the varying critical behavior near the first-order point where lattice effects are prominent. It is gratifying to note that our model Hamiltonian [Eq.~(\ref{H})] can capture a vast family of exponents in such systems, and RG values of $\beta$, $\gamma$, and $\delta$ are in excellent agreement with the experimental results. 

In the context of the strain-based theory parametrized by $\sigma$ presented here, it is germane to ask why $\sigma$ should change in manganites (R$_{1-x}$A$_x$MnO$_3$) on changing R, A and $x$. Different choices of $\rm R$ and $\rm A$ have different ionic radii and thus produce varying internal stresses on the  $\rm Mn$-$\rm O$-$\rm Mn$ bond length, changing the tolerance factor in Eq.~(\ref{f}). The bond stress results in electron transport being a sensitive function of the imposed strain due to perturbations induced via changes in R, A and $x$ \cite{am98,mdm98,bbk98,sxk96}. Thus, a change in composition
and doping in such systems produces varying strain interactions which spread over many lattice spacings. The dependence of $T_c$ on these strain modes
obeys a power law, as reported in Ref.~\cite{mdm98}. $T_c$ is also found to decrease with the doping parameter doping parameter $x$ \cite{rzb96}. Thus, we expect that $t$ and $x$ determine the strain exponent $\sigma$. The
precise relationship between $t$, $x$ and $\sigma$ is a theoretically challenging problem. Therefore, in this work we treat $\sigma$ as a parameter. As shown above, a reasonable range of $\sigma$-values dictated by RG enables us to capture the diverse critical behavior near and away from tricriticality in manganites.

\section{Summary and Discussion}
\label{s4}

Let us conclude this paper with a brief summary and discussion. We have
considered a model Hamiltonian $\mathcal H$ in Eq.~(\ref{H}), which is a functional of the spin and lattice degrees of freedom, with a coupling between them. The strain term is long-ranged and decays as a power-law with exponent $\sigma$. We have applied the Wilson RG scheme to this model, and derived the critical exponents corresponding to the nontrivial fixed point. The stability requirement restricts $\sigma$ to a range which depends on the dimension $d$ of the system. We find that $-1/2<\sigma<0$ for $d=3$, and $-1<\sigma<0$ for $d=2$. We have identified a small expansion parameter $\epsilon=4+2\sigma-d$, and obtained the critical exponents $\nu$ and $\eta$ to leading order in $\epsilon$. Using these two exponents, we obtained the other exponents $\alpha$, $\beta$, $\gamma$, and $\delta$ correct to $O(\epsilon)$.
  
In Table~\ref{comp}, we have compared our RG results with the available experimental exponents for manganite systems. We found that the RG values for $\beta$, $\gamma$, and $\delta$ are in excellent agreement with the experimental results. The tricritical exponents arise for the lower bound of $\sigma=-0.5$. As $\sigma$ increases towards the upper bound ($\sigma = 0$), the exponents move away from tricriticality. Thus, our theory (being an expansion about a tricritical point) yields exponents both near and far from tricriticality. It is gratifying to note that we can capture a large family of experimental results available for such systems. Clearly, our models yields a broad range of universality classes on varying $\sigma$.

The utility of this simple Hamiltonian in capturing the unconventional critical behavior of spin-lattice systems near their PM-FM phase transition is encouraging. The exotic behavior and phenomenology of such strongly correlated materials is due to complex interactions which couple spin and lattice degrees of freedom. The essential physics of CMR has long been assumed to be the interplay between a strong spin-lattice coupling and the spin-spin double-exchange effect. The double-exchange mechanism is widely regarded as the dominant physics, but it alone cannot account for the sharp resistivity changes near $T\sim T_c$ -- this requires a contribution from the electron-phonon coupling. A strong electron-phonon coupling may lead to the self-trapped state of an electron known as a polaron. Due to the tilting of Mn octahedra, the polaronic wave functions spread over many lattice sites, leading to the screening of these polaronic modes. However, it is still unknown whether the same Jahn-Teller-type electron-phonon interaction is at work in determining critical properties near the PM-FM phase transition. There are models based on Yukawa interactions \cite{hka06} showing that, for a much larger range of interactions, dimensionality $d$ plays a major role in determining the observed features in these compounds. Our RG results support this line of argument because the ensuing critical behavior is dictated by $d$ and $\sigma$. Since our theory captures the experimental results for infinite screening length, the long-ranged Fr{\"o}hlich-type electron-phonon interaction may play a crucial role in determining the PM-FM phase transition in such systems. However, further theoretical work is required to arrive at an unambiguous conclusion.

There has been a recent explosion of interest in complex functional materials, in which lattice distortions are coupled to electronic, magnetic and chemical degrees of freedom. This further emphasizes the need for a consistent theoretical framework to describe strain-based materials.
Manganites are very promising for developing advanced electronic devices \cite{ris06}. This is because, apart from exhibiting CMR (the dramatic response of resistivity to an external magnetic field), they also show {\it anisotropic magnetoresistance} or AMR (the response of resistivity to the field orientation vis-a-vis the crystal axis). The latter was shown \cite{lww09} to have a sensitive dependence on crystalline anisotropy, indicating a strong spin-lattice coupling. In addition to its direct relevance to the critical properties of manganites near the PM-FM phase transition, we believe our work is of broader significance in the critical theory of other strongly-correlated condensed matter systems.

An immediate extension of the present work is to study similar nonlocal quartic coupling in itinerant ferromagnets and antiferromagnets, and determine their anomalous scaling at quantum critical points. In itinerant electron systems, first-order quantum phase transitions have been the subject of extensive research \cite{ph02}. The Hertz-Millis-Moriya (HMM)\cite{am93} theory has been successfully applied to explain quantum critical behavior in a number of these materials. However, its assumption of analyticity of the quartic term has been questioned \cite{ac04}. It was shown that integrating the fermionic degrees of freedom in HMM theory leads to several discrepancies. Thus, to investigate the quantum critical point, the fermionic and bosonic degrees of freedom have to be treated on the same footing, and a nonlocal RG scheme similar to the one used in the present work may be appropriate. In addition, the critical dynamical behavior of many interacting systems, such as trapped ions, cavity quantum electrodynamics, Rydberg atom arrays, and cold atoms \cite{ddm22} also pose theoretical challenges where the nonlocal calculations of the present model could be useful.

\subsubsection*{Acknowledgments}

R.S. is grateful to the University Grants Commission, India for providing financial assistance through a D.S. Kothari postdoctoral fellowship. We are grateful to the referees for their constructive comments and suggestions.

\newpage
\hoffset-1in
\begin{table}[!]
{\caption{\small Critical exponents obtained from Eqs.~(\ref{betal})-(\ref{deltal}) in $d=3$ on varying the long-range exponent $\sigma \in$[-1/2,0]. RG exponents from the present theory are shown in brackets. We compare these with available experimental estimates on strain-coupled systems. Abbreviations: Polycrystalline (PC), Single-crystal (SC).}
\label{comp}}
\footnotesize
\begin{center}
\begin{tabular}{|c|c|c|c|c|c||}
\hline
$\sigma$  & Theory/ &  $\beta$ & $\gamma$ & $\delta$\\
& Experiments & & & \\
\hline
$-0.499$ &  Theory & 0.250 & 1.000 & 4.998 \\
\cline{2-5}
&  $\mbox{La}_{0.6}\mbox{Ca}_{0.4} \mbox{MnO}_3$ (PC)   \cite{krz02} & $0.25\pm0.03$ & $1.03\pm0.05$ & $5.0\pm0.8$\\
\cline{2-5}
&  $\mbox{La}_{0.6}\mbox{Ca}_{0.4} \mbox{MnO}_3$  (PC)  \cite{ntd13} & $0.248$ & $0.995$ & $4.896$\\
\cline{2-5}
&  $\mbox{Nd}_{0.67}\mbox{Sr}_{0.33} \mbox{MnO}_3$ (PC) \cite{vps08} & $0.23\pm0.02$ & $1.05\pm0.03$ & $5.13\pm0.04$\\
\cline{2-5}
&  $\mbox{La}_{0.1}\mbox{Nd}_{0.6}\mbox{Sr}_{0.3} \mbox{MnO}_3$ (PC) \cite{flh10} & $0.248\pm0.006$ & $1.066\pm0.002$ &--\\\hline

  $-0.466$ &  Theory & 0.257 & 1.013 & 4.941 \\\cline{2-5}
& $\mbox{La}_{0.1}\mbox{Nd}_{0.6}\mbox{Sr}_{0.3} \mbox{MnO}_3$ (PC) \cite{flh10} & $0.257\pm0.005$ & $1.12\pm0.03$ & $5.17\pm0.02$\\\hline

$-0.319$ & Theory & 0.288 & 1.076 & 4.613 \\\cline{2-5}
&  $\mbox{La}_{0.5}\mbox{Ca}_{0.3}\mbox{Ag}_{0.2} \mbox{MnO}_3$ (PC) \cite{swo14} & $0.288\pm0.002$ &$0.948\pm0.008$ &$4.90\pm0.02$\\\hline

$-0.290$ & Theory & 0.295 & 1.090 & 4.547\\\cline{2-5}
&  $\mbox{La}_{0.7}\mbox{Sr}_{0.3} \mbox{MnO}_3$ (SC) \cite{mse96} & $0.295\pm0.002$ &-- &--\\\hline

$-0.213$ & Theory & 0.314 & 1.128 & 4.380 \\\cline{2-5}
&  $\mbox{Pr}_{0.6}\mbox{Sr}_{0.4}\mbox{MnO}_3$ (PC) \cite{osh15} & $0.314\pm0.0006$ &$1.095\pm0.006$ &$4.545\pm0.008$\\\hline

$-0.194$ & Theory & 0.319 & 1.138 & 4.341 \\\cline{2-5}
&  $\mbox{La}_{0.825}\mbox{Sr}_{0.125}\mbox{MnO}_3$ (PC) \cite{tck20} & $0.319\pm0.001$ &$1.18\pm0.03$ &$4.67\pm0.03$\\\hline

$-0.188$ & Theory & 0.321 & 1.141 & 4.329 \\\cline{2-5}
&  $\mbox{Nd}_{0.6}\mbox{Sr}_{0.4}\mbox{MnO}_3$ (PC) \cite{osh15} & $0.321\pm0.003$ &$1.183\pm0.017$ &$4.75\pm0.02$\\\hline

$-0.176$ & Theory & 0.324 & 1.148 & 4.305 \\\cline{2-5}
&  $\mbox{La}_{0.67}\mbox{Sr}_{0.16}\mbox{Ca}_{0.17} \mbox{MnO}_3$ (PC) \cite{mtd15} & $0.324\pm0.005$ &$1.176\pm0.03$ &$4.415\pm0.02$\\\hline
	
$-0.161$ & Theory & 0.328 & 1.156 & 4.276 \\\cline{2-5}
&  $\mbox{La}_{0.8}\mbox{Ca}_{0.2} \mbox{MnO}_3$ \cite{ktb12} & $0.328$ &$1.193$ &$4.826$\\ \hline

$-0.103$ & Theory & 0.344 & 1.188 & 4.168 \\\cline{2-5}
&  $\mbox{Pr}_{0.77}\mbox{Pb}_{0.23} \mbox{MnO}_3$ (SC) \cite{pbe07} & $0.344\pm0.001$ & $1.352\pm0.006$ & $4.69\pm0.02$\\\hline

$-0.049$ & Theory & 0.360 & 1.220 & 4.077 \\\cline{2-5}
&  $\mbox{La}_{0.8}\mbox{Ca}_{0.2}\mbox{MnO}_3$ (SC) \cite{hkh01} & 0.36 & 1.45 & 5.03\\\hline

$-0.015$ & Theory & 0.370 & 1.241 & 4.023 \\\cline{2-5}
&  $\mbox{La}_{0.7}\mbox{Sr}_{0.3} \mbox{MnO}_3$ (SC) \cite{glg98} & $0.37\pm0.04$ & $1.22\pm0.03$ & $4.25\pm0.2$\\ \hline	
	
$-0.016$ & Theory & 0.370 & 1.240 & 4.024 \\\cline{2-5}
&  $\mbox{La}_{0.875}\mbox{Sr}_{0.125}\mbox{MnO}_3$ (SC) \cite{nbn03} & $0.37\pm0.02$  & $1.38\pm0.03$ & $4.72\pm0.04$\\\hline

$-0.003$ & Theory & 0.374 & 1.248 & 4.005\\\cline{2-5}
&  $\mbox{Nd}_{0.6}\mbox{Pb}_{0.4} \mbox{MnO}_3$ (SC) \cite{srg03} & $0.374\pm0.006$ & $1.329\pm0.003$ & $4.54\pm0.10$\\\hline
\end{tabular}
\end{center}
\end{table}

\newpage
\hoffset0in

\end{document}